\magnification=\magstep1
\input amstex
\documentstyle{amsppt}
\NoBlackBoxes

\hsize=5in
\vsize=7.5in

\rightheadtext{Classical $r$-Matrices and Compatible Poisson Structures}
\leftheadtext{Luen-Chau Li}

\def\pf{\hfill $\square$}
\def\c{\cite}
\def\M{\Cal M}
\def\RC{\Cal R}
\def\R{\Bbb R}
\def\Z{\Bbb Z}
\def\fr{\frac}
\def\const{\text{const}}
\def\kdv{\text{Korteweg de-Vries (KdV) }}
\def\tend{\text{End}}
\def\tinv{\text{inv}}
\def\tree{\text{tr}}
\def\men{\M_{\text{Benny}}}
\def\moda{\M_{d\text{Toda}}}
\def\g{\frak{g}}
\def\dl{\dot{L}}
\def\wtx{\widetilde{X}}
\def\wty{\widetilde{Y}}
\def\spaces{\;\big|\;}
\def\face{\{\cdot,\cdot\}}

\topmatter

\title
	Classical $\bold{r}$-Matrices and Compatible Poisson Structures
	for Lax Equations on Poisson Algebras
\endtitle

\author
	Luen-Chau Li   \\
\ \ \ \  \\
	{\rm Department of Mathematics  \\
	Pennsylvania State University  \\
	University Park, PA 16802 USA  \\
	luenli\@math.psu.edu}
\endauthor

\abstract
	Given a classical $r$-matrix on a Poisson algebra, we
	show how to construct a natural family of compatible
	Poisson structures for the Hamiltonian formulation of
	Lax equations.  Examples for which our formalism applies
	include the Benny hierachy, the dispersionless Toda
	lattice hierachy, the dispersionless KP and modified KP
	hierachies, the dispersionless Dym hierachy etc. 
\endabstract

\endtopmatter

\document
\baselineskip 20pt

\subhead
1. Introduction
\endsubhead

Two Poisson brackets on the same manifold are said to be compatible if
their sum is also a Poisson bracket \c{GDO, M}.  There are many
examples of integrable systems which are Hamiltonian with respect to
two compatible Poisson structures (see, e.g. \c{DO}).  Indeed, when
one of the structures happens to be nondegenerate, there is a simple
way which allows one to produce a whole family of compatible Poisson
structures \c{KR, RSTS1}.  However, the existence of further structures is
not a necessity when the two compatible structures are both
degenerate.

In the late seventies, we saw the beginning of the Lie algebraic
approach to integrable systems \c{K, A}.  The \kdv equation, for
example, was shown to be a Hamiltonian system on coadjoint orbits
\c{A}.  Furthermore, the second Poisson structure for KdV type
equations was constructed on subspaces of the algebra of formal
pseudo-differential operators \c{A, GD}.  We now refer to this
second structure as the Adler-Gelfand-Dickey structure.  Recently,
it was found to be of independent interest in conformal field
theory \c{DFIZ}.  In the mean time, the Lie algebraic approach 
to integrable systems was extensively developed, particularly by
the Russian school in St. Petersburg (see, e.g., the survey in
\c{RSTS2}).  In the so-called $r$-matrix framework, the simplest
Poisson structures for the Hamiltonian formulation of Lax equations on
Lie algebras are the linear Poisson structures associated with the
$R$-brackets.  In the case where $\g$ is the Lie algebra of a
noncommutative, associative algebra, a construction of quadratic
brackets which give Lax equations was first available for the
skew-symmetric $r$-matrices satisfying the modified Yang-Baxter
equation \c{STS1}.  Subsequently, this was superseded by a more general
construction valid for a wider class of $r$-matrices
\c{LP1,LP2}.  Indeed, in \c{LP2}, even a third order structure was 
found.  At this juncture, the reader should note that on the abstract
level of associative algebras, neither the linear structure nor the
quadratic structure is nondegenerate.  Therefore, the
recipe for producing a whole family of structures is not applicable
in this context.  As a matter of fact, no Poisson structures with order 
$> 3$ was ever found.  In this connection, we would like to mention 
the thesis of Strack \c{ST}, which showed (by using computer algebra)
that beyond order $3$, no Poisson structures of a certain form can 
exist for the Hamiltonian formulation of Lax equations.  So this is the
state of affairs for noncommutative, associative algebras.

In this paper, we address the Hamiltonian formulation of Lax
equations, as before, but in the context of Poisson algebras.  Here,
we show how to construct a natural family of compatible Poisson
structures on the full algebra.  On the group of invertible elements
(if non-empty and forms an open subset), similar consideration shows
we can even define structures of negative order.  Thus the situation
for Poisson algebras, in which multiplication is commutative, is
entirely different.  Recall that a Poisson algebra is by definition a
commutative, associative algebra with unit 1 equipped with a Lie
bracket such that the Leibniz rule holds \c{W1}.  The most familiar
examples of Poisson algebras are given by the collection of smooth
functions on Poisson manifolds.  For us, the particular examples which
have partly motivated this work are the algebras associated with the
truncated Benney's equation
\c{G-KR}, and the various dispersionless equations \c{DM, K, TT} which
are currently of interest in topological field theory \c{D, K}.  As
the reader will see, a family of vector fields $V_n$, $n \geq -1$,
plays the key role in this investigation.  These vector fields $V_n$
are invariants of degree 1 of the vector fields associated with the
Lax equations, and satisfy the Virasoro relations $[V_m,V_n] = (n - m)
V_{m+n}$.  For a given classical $r$-matrix on the Poisson algebra, we
can construct the associated linear bracket.  If we denote by
$\pi_{-1}$ the bivector field corresponding to this basic linear
structure, we shall show that the Lie derivatives $L_{V_m} \pi_{-1}$
essentially generate all higher order structures.  Thus our
construction works for an arbitrary classical $r$-matrix!  This is in
marked contrast to previous results on quadratic Poisson structures on
noncommutative, associative algebras \c{STS1, LP1, LP2}, where one has
to make rather stringent assumptions on the $r$-matrix.  In this
connection, we would like to remind the reader of the important
difference between the notions of double Lie algebras and Lie
bialgebras.  Recall that the former was motivated by the study of
integrable systems \c{STS1} and is associated with classical
$r$-matrices.  On the other hand, the notion of Lie bialgebras had its
origin in the geometry of Poisson Lie groups \c{DR}.  The two do
intersect, for example, in the class of double Lie algebras called
Baxter Lie algebras \c{STS2} (where the $r$-matrix satisfies
additional properties).  In our case, as the $r$-matrix is assumed to
be completely arbitrary, we are working within the framework of double
Lie algebras here.

The paper is organized as follows.  In sec. 2, we assemble a number of
basic facts and definitions which will be used in the paper.  In
Sec. 3, we formulate the main result and display the explicit formulas
for the linear, quadratic, and higher order structures.  Then we study
a number of basic properties.  In order to prepare for the proof of the
main result, we introduce the vector fields $V_n$ in Sect. 4 and
discuss their relation with the Lax equations.  Then, in Sect. 5,
we give a proof of the main result.  In order to illustrate the
use of our construction in Sect. 3, we describe the multi-Hamiltonian
formalism of some concrete partial differential equations in Sect. 6.
Our examples include the hierachy of truncated Benny equations
\c{B, G-KR} in nonlinear waves, the dispersionless Toda lattice
hierachy \c{DM}, the dispersionless KP \c{K, TT} and modified KP
hierachies, and the dispersionless Dym hierachy.  Note that in each
example, the set of Lax operators under consideration is a submanifold
of the full Poisson algebra.  However, this submanifold is not
necessarily a Poisson submanifold of the full algebra equipped with a
bracket which comes from Sect. 3.  For this reason, the passage from 
the bracket on the algebra to the Hamiltonian structure on the
submanifold of Lax operators might involve the process of reduction
\c{MR}.  Thus in our examples, we find Dirac reduction \c{D, MR}
(i.e. reduction with constraints) comes in naturally.  For the Benny
hierachy and the dispersionless Toda lattice hierachy, we shall
compute the first few Poisson structures explicitly, and illustrate
the use of Dirac reduction.  Our explicit expressions for the 
structures not only allow us to find the Casimir functions, they also
show that the structures which come from our Poisson algebras are of
hydrodynamic type or its generalizations \c{DN, F}.  Indeed, as it
turns out, all the higher structures of the dispersionless Toda
lattice hierachy are nonlocal generalizations of brackets of 
hydrodynamic type.  This shows how our construction in Sect. 3
can get complicated upon reduction to a specific submanifold of
Lax operators.

To close we stress again that our main result is formulated along
the lines of the $r$-matrix approach (where Poisson structures are
defined either on Lie algebras or their duals, or on Lie groups)
and applies to all Poisson algebras satisfying the assumptions
of Theorem 3.2.  In any concrete applications, the use of reduction
techniques (where necessary) is perfectly natural and the reader
should not feel uncomfortable under such circumstances.

\subhead
2. Preliminaries
\endsubhead

We collect in this section a number of basic facts, and introduce some
terminology which will be used in the sequel.

Let $P$ be a smooth manifold.
A Poisson bracket $\face$ on $P$ is a Lie bracket
on $C^{\infty}(P)$ which satisfies the derivation property in each
argument.  If $\pi$ is the bivector field corresponding to the bracket
operation, i.e.
$$
	\{F,H\} = \pi (dF,dH)\,,
\tag2.1
$$
then it is well-known that the Jacobi identity for $\face$
is equivalent to $[\pi,\pi]_S = 0$ \c{W2}, where $[\cdot,\cdot]_S$ is
the Schouten bracket $[S]$.  Recall that if $\Gamma(\wedge^k TM)$
is the space of sections of the vector bundle $\wedge^k TM$, and
$\wedge^*(M) = \underset{k\geq 0}\to\oplus \Gamma(\wedge^k TM)$,
the Schouten bracket $[\cdot,\cdot]_S$ is the bilinear map
$$
	[\cdot,\cdot]_S : \wedge^*(M) \times \wedge^* (M) \to 
	\wedge^*(M)
\tag2.2	
$$
which extends the usual Lie bracket operation on $\Gamma(TM)$ and makes
$\wedge^*(M)$ into a Lie superalgebra.  In particular, the following 
graded Jacobi identity holds:
$$
	(-1)^{pr} [u,[v,w]_S]_S + (-1)^{qp} [v,[w,u]_S]_S +
	(-1)^{rq} [w,[u,v]_S]_S = 0
\tag2.3
$$
where $u \in \Gamma(\wedge^p TM)$, $v \in \Gamma(\wedge^q TM)$ and
$w \in \Gamma(\wedge^r TM)$.  

As we mentioned in the introduction, two Poisson brackets on $P$ are
said to be compatible if their sum is also a Poisson bracket, i.e.
satisfies the Jacobi identity \c{GDO, M}.  In terms of the 
corresponding bivector fields $\pi_1$ and $\pi_2$, this is
equivalent to $[\pi_1,\pi_2]_S = 0$, as $[\pi_i,\pi_i]_S = 0$, 
$i = 1,2$. 

In this paper, we shall construct compatible Poisson structures for the
Hamiltonian formulation of Lax equations (associated with
$r$-matrices) when the underlying manifold $P$ is a Poisson algebra.

\definition 
{Definition 2.4}  Let $A$ be a commutative, associative algebra with
unit 1.  If there is a Lie bracket on $A$ such that for each element
$a \in A$, the operator $ad_a : b \mapsto [a,b]$ is a derivation of
the multiplication, then $(A, [\cdot,\cdot])$ is called a Poisson
algebra.
\enddefinition

Thus the Poisson algebras are Lie algebras with an additional 
associative algebra structure (with commutative multiplication and
unit 1) related by the derivation property to the Lie bracket.  Note
that some authors call the Lie bracket on $A$ the Poisson structure
on $A$ (see, for example, \c{W1}), but we shall refrain from such
usage in order to avoid confusion.

We now recall the notion of a classical $r$-matrix \c{STS1}.  Let
$\g$ be a Lie algebra.  A linear operator $R$  in the space $\g$ is
called a classical $r$-matrix if the $R$-bracket given by
$$
	[X,Y]_R = \fr12\; ([RX,Y] + [X,RY])\,, \qquad X,Y \in \g
\tag2.5
$$
is a Lie bracket, i.e.  satisifes the Jacobi identity.  Some
well-known sufficient conditions for $R \in \tend(\g)$ to be a
classical $r$-matrix are the Yang-Baxter equation and the modified
Yang-Baxter equation.  But in this paper, we can establish our results
without assuming these conditions.

To close this section, we define what we mean by Lax equations.

\definition
{Definition 2.6}  Let $A$ be a Poisson algebra, and suppose 
$R \in \tend(A)$ is a classical $r$-matrix.  Equations of the form
$$
	\dl = [R(X(L)),L]\,, \qquad L \in A
\tag2.7
$$
where $X : A \to A$ is a smooth map satisfying
$$
	[X(L),L] = 0\,, \qquad dX(L)\cdot [L',L] = [L',X(L)]\,,
	\qquad L , L' \in A
\tag2.8
$$
are called Lax equations.
\enddefinition

The basic Lax equations on $A$ are given by
$$
	\dl = Z_m(L) = [R(L^m),L], \qquad m \geq 1\,.
\tag2.9
$$
More generally, if $H$ is a smooth ad-invariant function (in the
sense defined in (3.1)), then $\dl = [R(L^m dH(L)), L]$ is also
a Lax equation.  

\subhead
3.  A Family of Compatible Poisson Structures on Poisson Algebras
\endsubhead

In what follows, we shall assume the Poisson algebra $A$ is
equipped with a non-degenerate ad-invariant pairing $(\cdot,\cdot)$.
A function $F$ defined on $A$ is said to be smooth if there exists
a map $dF : A \to A$ such that
$$
	\fr{d}{dt}\big|_{t = 0} F(L + t L') = (dF(L),L') \quad,
	\quad L, L' \in A
\tag3.1
$$

\proclaim
{Theorem 3.2}  Let $A$ be a Poisson algebra with Lie bracket
$[\cdot,\cdot]$ and non-degenerate ad-invariant pairing $(\cdot,\cdot)$
with respect to which the operation of multiplication is symmetric,
i.e. $(XY,Z) = (X,YZ)$, $ \forall\,X,Y,Z \in A$.  Assume $R \in \tend(A)$
is a classical $r$-matrix, then 
\roster
\item"{(a) }"  for each integer $n \geq - 1$, the formula
$$
	\{F,H\}_{(n)}(L) = (L,[R(L^{n+1} dF(L)), dH(L)] + [dF(L),
	R(L^{n+1} dH(L))])
\tag3.3
$$
(where $F$ and $H$ are smooth) defines a Poisson structure on $A$, 
\item"{(b) }" the structures $\face_{(n)}$ are compatible
	with each other,
\item"{(c) }"  if $\pi_n$ is the bivector field corresponding to 
$\face_{(n)}$ and $D_{\pi_n} : \wedge^*(A) \to \wedge^*(A)$ 
is the associated coboundary operator, i.e. $D_{\pi_n} X = [ \pi_n,X]_S$,
$X \in \wedge^* (A)$.      There exists vector fields $V_m$ on $A$,
$m \geq -1$ satisfying the Virasoro relations $[V_m,V_n] = (n - m)
V_{m+n}$ such that $D_{\pi_n} V_m = (n - m) \pi_{m+n}$, $m,n \geq - 1$.
\endroster
\pf
\endproclaim

We shall prove this result in Section 5, after we introduce the 
vector fields $V_m$ in Section 4 and explain what they are in relation
to the Lax equations.  As the reader will see, the relations
$[\pi_n,V_m]_S = (n - m) \pi_{m+n}$ between the bivector fields
which we establish at the beginning of Section 5 play the key role
in proving parts (a) and (b) of the above theorem.  They are also
responsible for the following.

\proclaim
{Corollary 3.4} \ (Involution of Casimir Functions) $\{H_{\pi_n}^0(A),
H_{\pi_n}^0(A)\}_{(m+n)} = 0$, $m,n \geq - 1$.  $m \neq n$.
\endproclaim  

\demo
{Proof} This follows from the formula $[\pi_n,V_m]_S (dF,dH) = L_{V_m}
\pi_n (dF,dH) = \mathbreak V_m \{F,H\}_{(n)} - \{V_m F,H\}_{(n)} - \{F,V_m
H\}_{(n)}$.  \pf 
\enddemo 

\remark
{Remark 3.5}  Note that from the compatibility of the structures, it
follows that
$$
	\{H_{\pi_m}^0(A),H_{\pi_m}^0(A)\}_{(n)} \subset H_{\pi_m}^0(A)\,.
\tag3.6
$$
\endremark

We now give a number of basic properties of the Poisson structures
$\face_{(n)}$, $n \geq - 1$.

\proclaim
{Theorem 3.7}  (a) \ Smooth functions in $A$ which are ad-invariant
Poisson commute in $\face_{(n)}$.

(b) \ The Hamiltonian system generated by a smooth ad-invariant
function $H$ in the Poisson structure $\face_{(n)}$
is given by the Lax equation $\dl = [R(L^{n+1} dH(L)),L]$.
\endproclaim

\demo
{Proof} (a) \ If $F$ and $H$ are smooth functions in $A$ which are
ad-invariant, we have $[dF(L),L] = [dH(L),L] = 0$.  Therefore,
$\{F,H\}_{(n)}(L) = ([dH(L),L],R(L^{n+1} dF(L))) + ([L,dF(L)],
\;R(L^{n+1} dH(L))) = 0$.

(b) \ If $H$ is ad-invariant, for any smooth $F$, we have 
$\{F,H\}_{(n)}(L) = (L,[dF(L)$, $R(L^{n+1} dH(L))]) = (dF(L),\;
[R(L^{n+1} dH(L)),L])$.  \pf
\enddemo

>From formula (3.3), it is clear that the bracket $\face_{(n)}$
vanishes at the unit 1.  Therefore, the linearization of $\{\cdot,
\cdot\}_{(n)}$ defines a Lie bracket on $A$, and an easy calculation
shows it coincides with the  $R$-bracket $[\cdot,\cdot]_{R}$.

The following result is reminiscent of the multiplicative property
of Poisson Lie groups \c{DR}.  However, it is in the context of
a Poisson algebra and the reason for its validity is entirely
different.

\proclaim
{Theorem 3.8}  Equip $A$ with the structure $\face_{(0)}$
and $A \times A$ with the product structure.  Then the multiplication 
map $m : A \times A \to A$ is a Poisson map.  
\endproclaim

\demo
{Proof}
Let $F$ and $H$ be smooth functions on $A$.  For $L_1, L_2 \in A$,
let $L = m(L_1,L_2)$.  Clearly, $F \circ m$ depends on two variables
and by taking its derivative with respect to the $i$-th variable,
$i = 1,2$, we obtain $d_1(F \circ m) (L_1,L_2) = L_2 d F(L)$,
$d_2 (F \circ m)(L_1, L_2) = L_1 d F(L)$.  To simplify notation,
let $X_1 = d F(L)$, $X_2 = d H(L)$ and denote the product structure
on $A \times A$ also by $\face_{(0)}$, then we have
$$
\align
	& \{ F \circ m,H \circ m\}_{(0)} (L_1,L_2)   \tag"{(*)}"
	\\
	& = (L_1, [R(LX_1),L_2 X_2] + [L_2 X_1, R(LX_2)])
	\\
	& + (L_2, [R(LX_1),L_1 X_2] + [L_1 X_1, R(LX_2)])\,.
\endalign
$$
By the derivation property of $[\cdot,\cdot]$, the commutativity
of multiplication and its symmetry with respect to the ad-invariant
pairing $(\cdot,\cdot)$, we have
$$
\aligned
	& (L_1, [R(LX_1),L_2 X_2])  \\
	& = (L,[R(LX_1),X_2]) - (L_2, [R(LX_1),L_1 X_2])\,.
\endaligned
$$
Likewise,
$$
\aligned
	& (L_2, [L_1 X_1,R (LX_2)])  \\
	& = (L,[X_1, R (LX_2)] - (L_1, [L_2 X_1, R (LX_2)])\,.
\endaligned
$$
When we insert these relations in (*), the result follows.
\pf
\enddemo

Consider now $A_{\tinv}$, the group of invertible elements of $A$.
We assume $A_{\tinv} \neq \phi$ and form an open subset of $A$.
Then we can define vector fields $Z_{-m}$, $V_{-n}$ for $m \geq 1$,
$n \geq 2$, on $A_{\tinv}$ as in formulas (4.2) and (4.5).  If we
define 
$$
	\{F,H\}_{(-n)}(L) = (L,[R(L^{-n+1}dF(L)),dH(L)] +
	[dF(L),R(L^{-n+1}dH(L))]), \quad n \geq 2
\tag3.9
$$
for smooth functions $F$ and $H$ on $A_{\tinv}$, it is easy to check
that the analysis in Section 5 also holds for these objects.  In
particular, this means $\face_{(-n)}$ are Poisson structures
on $A_{\tinv}$.

\proclaim
{Theorem 3.10} Let $\iota: A_{\tinv} \to A_{\tinv}$ be the inversion
map, i.e. $\iota(L) = L^{-1}$.  Then $\{F \circ \iota, H\circ
\iota\}_{(n)} (L) = -\{F,H\}_{(-n)} \circ \iota(L)$, $n \geq 0$, for
all smooth functions $F$ and $H$ on $A_{\tinv}$.
\endproclaim

\demo
{Proof}  We have $d(F \circ \iota)(L) = 
- L^{-2} dF(L^{-1})$ and so \newline
$\{F \circ \iota,H\circ \iota\}_{(n)} (L) = (L,[R(L^{n-1}dF(L^{-1})),
\;L^{-2}dH(L^{-1})] - (F \leftrightarrow H))$.  
Now,
$$
\aligned
	& (L,\;[R(L^{n-1}dF(L^{-1})),L^{-2}dH(L^{-1})])    \\
	= & (L,\;L^{-2}[R(L^{n-1}dF(L^{-1})),dH(L^{-1})]) + (L\,dH(L^{-1}),
		\;[R(L^{n-1} dF(L^{-1})),\;L^{-2}])   \\
	= & (L^{-1},\;[R(L^{n-1}dF(L^{-1})),\;dH(L^{-1})]) + 2
		(dH(L^{-1}),\;[R(L^{n-1} dF(L^{-1})),\;L^{-1}])  \\
	= & - (L^{-1}, [R(L^{n-1} dF(L^{-1})),\;dH(L^{-1})])\,.
\endaligned
$$
Hence the assertion follows.  \pf
\enddemo

\subhead
4. \ Lax Equations on Poisson Algebras and Virasoro Invariants
\endsubhead

According to Definition 2.6, corresponding to each smooth map 
$X : A \to A$ satisfying (2.8) is a Lax equation
$$
	\dl = \wtx(L) = [R(X(L)),L]
\tag4.1
$$
To prepare for the proof of Theorem 3.2, we shall introduce vector
fields $V_n$, $n \geq -1$ on $A$ which are related to the Lax
equations.  Before we do so, we first prove

\proclaim
{Theorem 4.2}  Let $X,Y : A \to A$ be smooth maps satisfying (2.8).
Then $[\wtx,\wty] = 0$.
\endproclaim

\demo
{Proof}  We have
$$
\aligned
	& d\wty(L)\cdot \wtx(L)   \\
	= & [R(dY(L)\cdot\wtx(L)),L] + [R(Y(L)),\;\wtx(L)]    \\
	= & [R([R(X(L)),\;Y(L)]),L] + [R(Y(L)),\; [R(X(L)), L]]\,.
\endaligned	
$$
Therefore,      
$$
\aligned
	& [\wtx,\wty](L)   \\
	= & 2[R([X(L),Y(L)]_R),\;L] + [R(Y(L)),\;[R(X(L)),L]] -
		[R(X(L)),\;[R(Y(L)),L]]    \\
	= & [2R([X(L),Y(L)]_R),\;L] - [[R(X(L)),\;R(Y(L))],\;L]\,,
		\text{ by Jacobi identity}   \\
	= & - [[R(X(L)),\;R(Y(L))] - 2R([X(L),Y(L)]_R),\;L]\,.
\endaligned
$$
Let $B_R(X,Y) = [RX,RY] - 2R([X,Y]_R)$.  Then $R$ is a classical
$r$-matrix iff $[B_R(X,Y),Z] + [B_R(Y,Z),X] + [B_R(Z,X),Y] =
0$, $\forall\;X,Y,Z \in A$.  Using the ad-invariant pairing, this
is equivalent to         
$$
\aligned
	[B_R(X,Y),Z] = & R^*[RX,[Y,Z]] - R^*[X,R^*[Y,Z]] - [RX,R^*[Y,Z]]
	\\
	& + R^*[RY,[Z,X]] - R^*[Y,R^*[Z,X]] - [RY,R^*[Z,X]]
\endaligned
$$
If we now put $X = X(L)$, $Y = Y(L)$ and $Z = L$ in the above
relation, we obtain $[\wtx,\wty](L) = 0$, as asserted.  \pf
\enddemo

The vector fields $V_n$, $n \geq - 1$, are defined as follows:
$$
	V_n(L) = L^{n+1}\,, \qquad n \geq - 1
\tag4.3
$$

\proclaim
{Theorem 4.4}  The vector fields $V_n$ satisfy the Virasoro
relations $[V_m,V_n] = (n - m) V_{m+n}$, $m,n \geq - 1$.
\endproclaim

\demo
{Proof}  Clear.  \pf 
\enddemo

Given a smooth manifold $M$ and a vector field $V$ on $M$,
recall that a tensor field $T$ is an invariant tensor field
of $V$ iff $L_V T = 0$.  Generalizing one step further, we
shall say that $T$ is an invariant tensor field of degree 1
iff $L_X^2 T = 0$.  The vector fields $V_n$ introduced in 
(4.3) above are invariants of degree 1 of the vector fields
$\wtx$ corresponding to the Lax equations.  Indeed, we have

\proclaim
{Theorem 4.5}  If $X : A \to A$ is a smooth map satisfying (2.8),
we have $L_{V_m} \wtx = \wty$ where $Y(L) = dX(L)\cdot V_m(L)$.
\endproclaim

\demo
{Proof}  
$$
\aligned
	& [V_m,\wtx] (L)  \\
	= & d\wtx(L) \cdot V_m(L) - dV_m(L),\wtx(L)   \\
	= & [R(dX(L) \cdot V_m(L)),L] + [R(X(L)),V_m(L)] - (m + 1) L^m
		[R(X(L)),L]  \\
	= & [R(dX(L),V_m(L)),L]\,.
\endaligned
$$
Thus, it remains to show $Y(L) = dX(L)\cdot V_m(L)$ satisfies (2.8).
To do this, first note that from the condition $[X(L),L] = 0$,
$L \in A$, we have $[dX(L)\cdot L',L] + [X(L),L'] = 0$, $L,L' \in A$.
Therefore, $[Y(L),L] = [dX(L) \cdot V_m(L),L] = - [X(L),V_m(L)] = - 
(m + 1)L^m [X(L),L] = 0$, for all $L \in A$.  On the other hand,
it follows from $dX(L) \cdot [L',L] = [L',X(L)]$, $L,L' \in A$,
that
$$
	(d^2 X(L)\cdot L')([L'',L]) + dX(L)\cdot [L'',L'] = [L'',dX(L)
	\cdot L'], L,L',L'' \in A\,.
\tag"{(*)}"
$$
Consequently, for all $L,L' \in A$, we have
$$
\aligned
	& dY(L) \cdot [L',L] = (d^2 X(L) \cdot [L',L])(V_m(L)) + 
	dX(L) \cdot ((m + 1) L^m [L',L])   \\
	& = (d^2 X(L) \cdot V_m(L))([L',L]) + dX(L)\cdot[L',V_m(L)]  \\
	& = [L',Y(L)], \text{ by (*)}.  
\endaligned
$$
\pf
\enddemo

\remark
{Remark 4.6}  For the vector fields $Z_n$ in (2.9), we have in 
particular the relations $L_{V_m} Z_n = n Z_{m + n}$, $m \geq -1$, 
$n \geq 1$.
\endremark

If we now combine Theorem 4.5 and Theorem 4.2, the nature of the
vector fields $V_m$ is now revealed.

\proclaim
{Corollary 4.7}  $L_{\wtx}^2 V_n = 0$, $n \geq 0$, $L_{\wtx} V_{-1}
= 0$.  \pf 
\endproclaim

\subhead
5. \ Virasoro Action on the Bivector Fields and Compatibility of 
the Structures
\endsubhead

The goal of this section is to prove Theorem 3.2.  To do this, we
consider the action of the vector fields $V_m$ on the bivector fields
$\pi_n$ corresponding to $\face_{(n)}$, $n \geq -1$.

\proclaim
{Theorem 5.1}  $L_{V_m} \pi_n = (n - m) \pi_{m+n}$, $m,n \geq - 1$.
\endproclaim

As indicated in Section 3, this result is the key in proving
Theorem 3.2.  The demonstration of Theorem 5.1 is quite tedious,
so we break it up into several steps.  First, note that from the
property of the Lie derivative, we have
$$
	L_{V_m} \pi_n (dF,dH) = V_m \{F,H\}_{(n)} - \{V_m F,H\}_{(n)}
	- \{F, V_m H\}_{(n)}\,.
\tag5.2
$$
Using the expressions for $\face_{(n)}$ and $V_m$, we 
obtain the identities in the next two lemmas.  We shall omit the
rather lengthy computations.  

\proclaim
{Lemma 5.3}  
$$
\aligned
	& V_m\{F,H\}_{(n)}(L) = (V_m(L),[R(L^{n+1} dF(L)),dH(L)])  \\
	& + (L,[R(L^{n+1}dF(L)),d^2 H(L) \cdot V_m(L)]) + (n + 1)(L,
		[R(L^{m+n+1}dF(L)),dH(L)])  \\
	& + (L,[R(L^{n+1} d^2 F(L) \cdot V_m(L)),dH(L)]) - (F 
		\leftrightarrow	H)
\endaligned
$$ 
where $(F \leftrightarrow H)$ denote terms obtained from previous
ones by switching $F$ and $H$.  
\line{\hfill\pf}
\endproclaim

\proclaim
{Lemma 5.4}  
$$
\aligned
	& \{V_m F,H\}_{(n)}(L) + \{F,V_m H\}_{(n)} (L) = (L,
		[R(L^{n+1} d^2 F(L) \cdot V_m(L))\,,   \\
	& dH(L)] + [d^2 F(L) \cdot V_m(L),R(L^{n+1}dH(L))]) + (m + 1)
		(L,[R(L^{m+n+1}dF(L)),dH(L)]  \\
	& + [L^m dF(L),R(L^{n+1}dH(L))]) - (F \leftrightarrow H)\,. 
\endaligned
$$
\pf
\endproclaim

\demo
{Proof of Theorem 5.1}  By combining the expressions in Lemma 5.3 
and Lemma 5.4 according to (5.2), it is clear that terms involving
second derivatives cancel out, and we obtain 
$$
\align
	& L_{V_m} \pi_n (L) \;(X_1,X_2)    \tag"{(*)}"   \\
	& = (V_m(L),[R(L^{n+1} X_1),X_2]) + (n - m) (L,[R(L^{m+n+1}
		X_1),X_2])   \\
	& - (m + 1)(L,[L^m X_1,R(L^{n+1} X_2)]) - (1 \leftrightarrow
		2)
\endalign
$$
where $X_1 = dF(L),X_2 = dH(L)$.  Now, by repeated application of
the derivation property, the commutativity of multiplication and 
its symmetry with respect to $(\cdot,\cdot)$, we have
$$
\aligned
	&(V_m(L), [R(L^{n+1}X_1),X_2]) - (1 \leftrightarrow 2)  \\
	& = (L,[R(L^{n+1}X_1),L^m X_2]) - (LX_2, [R(L^{n+1} X_1),L^m])
		- (1 \leftrightarrow 2)   \\
	& = (L,[R(L^{n+1}X_1),L^m X_2]) - m(L^m X_2,[R(L^{n+1}X_1),
		L]) - (1 \leftrightarrow 2)   \\
	& = (m + 1) (L,[R(L^{n+1}X_1),L^m X_2]) - (1 \leftrightarrow 2)
	\\
	& = (m + 1) (L,[L^{m}X_1, R(L^{n+1} X_2)]) - (1 \leftrightarrow 2)\,.
\endaligned
$$
If we substitute this in (*), the result follows.  \pf
\enddemo

\remark
{Remark 5.5}
In the case of noncommutative, associative algebra, relations
similar to the ones in Theorem 5.1 were obtained in \c{LP2}
for the three structures there.
\endremark

\proclaim
{Corollary 5.6}   $[\pi_m,\pi_n]_S = \fr1{n+2} [V_{n+1},[\pi_{-1},
\pi_m]_S]_S + \fr{m - n - 1}{n + 2} [\pi_{-1},\pi_{m+n+1}]_S$ 
for $m,n \geq -1$.
\endproclaim

\demo
{Proof}  From Theorem 5.1 and the graded Jacobi identity for the
Schouten bracket, it follows that
$$
\aligned
	& [\pi_m, \pi_n]_S   \\
	= & - \fr1{n+2}\; [\pi_m, [V_{n+1},\pi_{-1}]_S]_S   \\
	= & - \fr1{n+2}\; [V_{n+1},[\pi_{-1},\pi_m]_S]_S + \fr1{n+2}\;
		[\pi_{-1}, [V_{n+1},\pi_m]_S]_S   \\
	= & - \fr1{n+2}\; [V_{n+1},[\pi_{-1},\pi_m]_S]_S + \fr{m-n-1}{n+2}\;
		[\pi_{-1},\pi_{m+n+1}]_S
\endaligned
$$
\pf
\enddemo

\remark
{Remark 5.7}  The formulation of Corollary 5.6 is motivated by
similar considerations in \c{AvM}.
\endremark

\demo
{Proof of Theorem 3.2} If we set $m = -1$ in the identity in Corollary
5.5, we find $[\pi_{-1},\pi_n]_S = - \fr1{2(n+2)}\;
[V_{n+1},[\pi_{-1},\pi_{-1}]_S]_S = 0$, $\forall \;n \geq - 1$, as
$\pi_{-1}$ is the bivector field for the Lie-Poisson structure
$\face_{(-1)}$.  From the same identity, it now follows that
$[\pi_m,\pi_n]_S = 0$, $\forall\;m,n \geq -1$.  Hence the brackets
$\face_{(n)}$ define compatible Poisson structures on $A$.
Finally, the assertion in part (c) follows from Theorem 5.1.  \pf
\enddemo

\subhead
6. \ Some Examples
\endsubhead

In this section, we look at some concrete examples of partial
differential equations which can be realized as Lax equations 
on Poisson algebras.  In each case, we describe the multi-Hamiltonian
formalism which follows from our universal construction in Sect. 3.
The reader should note that in these applications, we are dealing with
Lax operators which form submanifolds of the full Poisson subalgebras
under consideration.  Although these submanifolds of Lax operators
are invariant under the dynamics of the associated Lax equations,
however, they are not automatically Poisson submanifolds of the
brackets which arise from the general construction in Sect. 3.
For this reason, there are two kinds of situations in the
examples which follows.  In the happy case where the submanifold
$\M$ of Lax operators does form a Poisson submanifold of $(A,
\face_{(n)})$, there is of course an induced structure on $\M$
which can be obtained by simple restriction of $\{\cdot,\cdot\}_{(n)}$
to $\M$.  On the other hand, when $\M$ is not a Poisson submanifold
of $(A,\{\cdot,\cdot\}_{(n)})$, the reader will see that the geometry
in each case warrants the application of Dirac reduction, i.e.
reduction with constraints \c{D, KO, MR}.  Thus in this latter case,
the brackets which arise from the construction in Sect. 3 serve
as the starting point of a reduction process from which the 
constrained brackets on $\M$ are computed.

In the following, we shall rescale the expression for $\{\cdot,
\cdot\}_{(n)}$ by the factor $\fr12$.  

\subhead
$1^{\circ}$ \ The Benny hierachy
\endsubhead

The Benny equations in nonlinear waves \c{B} (we shall consider
the simplest case here) are given by the quasi-linear system
$$
	\pmatrix u_0 \\ u_{-1} \endpmatrix_t = 
	\pmatrix u_0 & 1 \\ u_{-1} & u_0 \endpmatrix 
	\pmatrix u_0 \\ u_{-1} \endpmatrix_x
\tag6.1
$$
We shall deal with the case where $u_0,u_{-1}$ are smooth 
functions on the circle $S^1 = \R/\Z$.

Following \c{G-KR},  introduce the algebra $A$ of Laurent polynomials
in $\lambda$, having the form
$$
	u(x,\lambda) = \sum_i u_i(x)\lambda^i\,,
\tag6.2
$$
where the coefficients $u_i$ are smooth functions on the circle
$S^1$.  With the well-known Lie-bracket defined by
$$
	[u,v]_{-1} = \fr{\partial u}{\partial\lambda}\;
	\fr{\partial v}{\partial x} - \fr{\partial u}{\partial x}\;
	\fr{\partial v}{\partial\lambda}\;, \qquad
	u,v \in A\,,
\tag6.3
$$
it is clear that $(A,\;[\cdot,\cdot]_{-1})$ is a Poisson algebra.
In \c{G-KR}, the Benny equations are rewritten as a Lax equation
in this Poisson algebra.  Indeed, (6.1) is equivalent to
$$
	\fr{dL}{dt} = \left[R\left(\fr14 L^2\right)\,, L\,\right]_{-1}
\tag6.4
$$
where the Lax operator $L$ is an element of the Benny manifold
$$
	\men = \left\{ L \in A \spaces L(x,\lambda) =
	\lambda + u_0(x) + u_{-1}(x)\lambda^{-1}\right\}
\tag6.5
$$
and the $r$-matrix $R$ is the one associated with the direct
sum decomposition
$$
	A = A_{\geqslant 1} \oplus A_{\leqslant 0}
\tag6.6
$$
into subalgebras
$$
	A_{\geqslant 1} = \left\{ u \in A \spaces u(x,\lambda)
	= \sum_{i\geqslant 1} u_i(x)\lambda^i \right\}
\tag6.7a
$$
$$
	A_{\leqslant 0} = \left\{ u \in A \spaces u(x,\lambda)
	= \sum_{i\leqslant 0} u_i(x)\lambda^i \right\}\;.
\tag6.7b
$$
In view of the representation in (6.4), the quasi-linear system
(6.1) is only a member of a hierachy of Lax equations on 
$\men$, and this is what we call the Benny hierachy.
Note that the Poisson algebra introduced above admits the
trace functional
$$
	\tree_{-1} u = \int u_{-1}(x) dx\,, \qquad u \in A
\tag6.8
$$
(here and below we integrate over $S^1$) \newline
which satisfies the important property
$$
	\tree_{-1}[u,v] = 0\,, \qquad u,v \in A\,.
\tag6.9
$$
Therefore, we can equip $A$ with a non-degenerate ad-invariant
pairing $(\cdot,\cdot)_{-1}$:
$$
	(u,v)_{-1} = \tree_{-1}(uv)\quad,\qquad u,v \in A\,.
\tag6.10
$$
Thus we have all the ingredients which are required for the
application of Theorem 3.2.  Consequently, we have a family
of Poisson structures $\{\cdot,\cdot\}_{(n)}$, $n \geq -1$,
on $A$.

It is easy to check that $\men$ is a Poisson 
submanifold of $(A,\face_{(-1)})$.  Therefore, the
induced structure on $\men$ provides the first 
Poisson structure for the equations in the Benny hierachy
\c{G-KR}.  Using $u = (u_0,u_{-1})$ as coordinates on
$\men$, the associated Hamiltonian operator is given 
explicitly by 
$$
	B_{(-1)}(u) = \pmatrix 0 & D \\ D & 0 \endpmatrix
	\;,\qquad D = \fr{d}{dx}
\tag6.11
$$
which is apparently well-known to people working in other
frameworks (see, for example, \c{DN}).  Clearly, this first
structure is degenerate, with Casimirs given by
$C_1(u) = \int u_0(x) dx$ and $C_2(u) = \int u_{-1}(x) dx$.

\remark
{Remark 6.12}  One of the advantages in formulating the Benny
equations as a Lax equation on $A$ is that it automatically
suggests a method of solution, namely, via a factorization 
problem on a symplectic diffeomorphism group.  The analytic
details, however, are nontrivial.
\endremark

We now turn to the higher structures.  Here, it is easy to see that
$\men$ is not a Poisson submanifold of any of the brackets
$\face_{(n)}$, $n\geqslant 0$.  However, we shall see
that we can apply Dirac reduction to $\face_{(n)}$ 
with appropriate constraints to obtain the higher structures
on $\men$.  We shall illustrate the procedure for $n = 0$ and
$n = 1$, thereby obtaining the second and third Poisson 
structures on $\men$.

For $n = 0$, the Hamiltonian vector field generated by $H$ is
of the form
$$
\align
	X_H^{(0)} (u) & = u \Pi_{\leqslant - 2} ([dH(u),u]_{-1})
	- \left[ \Pi_{\leqslant 0} (udH(u)), u\right]_{-1}
	\tag6.13    \\
	& = \left[\Pi_{\geqslant 1} (udH(u)),u\right]_{-1} 
	- u \Pi_{\geq - 1} ([dH(u),u]_{-1})\,.
\endalign
$$
If $L \in \men$, it follows from this formula that the highest
order term of $X_H^{(0)}(L)$ in $\lambda$ is $\lambda^0$, while
the lowest order is in $\lambda^{-2}$.  Using $u = (u_0,u_{-1},u_{-2})$
as coordinates on the submanifold $\{\lambda + u_0(x) + u_{-1}(x)
\lambda^{-1} + u_{-2}(x) \lambda^{-2} \in A\}$, the operator
which gives $X_H^{(0)}(L)$ can be computed explicitly:
$$
	\pmatrix
		D & u_0 D + u_{0x} & u_{-1}D + u_{-1x}  \\
		u_0 D & 2u_{-1} D + u_{-1x} & 0  \\
		u_{-1} D & 0 & -u_{-1}^2 D - u_{-1} u_{-1x}
	\endpmatrix
\tag6.14
$$
Therefore, we can apply Dirac reduction with constraint
$u_{-2} \equiv 0$ to obtain the second structure on $\men$:
$$
\align
	B_0(u) & = \pmatrix D & u_0 D + u_{0x} \\ u_0 D &
		2u_{-1}D + u_{-1x} \endpmatrix - \pmatrix
		u_{-1} D + u_{-1x}  \\ 0 \endpmatrix 
		(-u_{-1}^2 D - u_{-1} u_{-1x})^{-1} (u_{-1} D\;0)
	\tag6.15
	\\
	& = \pmatrix 2 & u_0 \\ u_0 & 2u_{-1} \endpmatrix D 
		+ \pmatrix 0 & 1 \\ 0 & 0 \endpmatrix u_{0x}
		+ \pmatrix 0 & 0 \\ 1 & 0 \endpmatrix u_{-1x}
\endalign
$$
Note that this second structure is of hydrodynamic type \c{DN}
because the associated Hamiltonian operator is of the form
$$
	B_0^{ij}(u) = g^{ij}(u) D + b_k^{ij}(u) \;u_x^k\,.
\tag6.16
$$
In this case, the metric which defines the structure (6.15)
is non-degenerate where
$$
	\Delta = u_0^2 - 4u_{-1} \neq 0\,.
\tag6.17
$$

For $n = 1$, i.e. for the bracket $\face_{(1)}$, we
have a similar formula for the Hamiltonian vector field
$$
\align
	X_H^{(1)}(u) & = u^2 \Pi_{\leqslant - 2}([dH(u),u]_{-1})
		- \left[ \Pi_{\leqslant 0} (u^2 dH(u)),u\right]_{-1}
		\tag6.18
	\\
	& = \left[\Pi_{\geqslant 1} (u^2 dH(u)),u\right]_{-1}
		- u^2 \Pi_{\geqslant - 1} ([dH(u),u]_{-1})\,.
\endalign
$$
This time, the highest order term of $X_H^{(1)}(L)$ $(L \in\men)$
in $\lambda$ is still $\lambda^0$, but the lowest order term is in 
$\lambda^{-3}$.  Therefore, in the coordinates $u = (u_0,u_{-1},
u_{-2},u_{-3})$, the operator which gives $X_H^{(1)}(L)$ is 
given by
$$
\left(
	\smallmatrix
	2u_0 D + u_{0x} & (u_0^2 + 3u_{-1})D + 2u_0 u_{0x} + 2 u_{-1x}
	& 2u_0 u_{-1} D + 2u_0 u_{-1x} + 2u_{-1} u_{0x} & u_{-1}^2
	D + 2u_{-1} u_{-1x}
	\\
	(u_0^2 + 3u_{-1})D + u_{-1x} & 4u_0 u_{-1}D + 12 u_{-1}u_{0x}
	+ 2u_0 u_{-1x} & u_{-1}^2 D + 2u_{-1} u_{-1x} & 0
	\\
	2u_0 u_{-1} D & u_{-1}^2 D & -2 u_0 u_{-1}^2 D - 2u_0 u_{-1} u_{-1x}
	- u_{-1}^2 u_{0x} & -u_{-1}^3 D - 2u_{-1}^2 u_{-1x}
	\\
	u_{-1}^2 D & 0 & -u_{-1}^3 D - u_{-1}^2 u_{-1x} & 0
	\endsmallmatrix
\right)
\tag6.19
$$
\vskip .05in
\noindent To obtain the structure on $\men$, we have to use Dirac 
reduction with the constraints $u_{-2} \equiv 0$, $u_{-3} \equiv 0$.
Accordingly, we have to invert the lower $2 \times 2$ block of (6.19):
$$
\gathered
	\pmatrix
	-2 u_0 u_{-1}^2 D - 2u_0 u_{-1} u_{-1x} - u_{-1}^2 u_{0x}
	& -u_{-1}^3 D - 2u_{-1}^2 u_{-1x} 
	\\
	-u_{-1}^3 D - u_{-1}^2 u_{-1x} & 0
	\endpmatrix^{-1}  
	\\
	= - \pmatrix
	0 & \fr1{u_{-1}^2} D^{-1} \fr1{u_{-1}^2} 
	\\
	\fr1{u_{-1}^2} D^{-1} \fr1{u_{-1}}
	&
	-\fr1{u_{-1}^2} D^{-1} \fr{u_0}{u_{-1}^2} -
	\fr{u_0}{u_{-1}^2} D^{-1} \fr1{u_{-1}^2}
	\endpmatrix
\endgathered
\tag6.20
$$
Hence the Hamiltonian operator of the third structure is given
by
$$
\align
	B_1(u) & = \left(\smallmatrix
	2u_0 D + u_{0x} & (u_0^2 + 3u_{-1})D + 2u_0 u_{0x} +
	2u_{-1x}   \\
	(u_0^2 + 3u_{-1}) D + u_{-1x} & 4 u_0  u_{-1} D + 2u_{-1}
	u_{0x} + 2u_0 u_{-1x}
	\endsmallmatrix\right)  \tag6.21
	\\
	& + \left(\smallmatrix
	2u_0 u_{-1} D + 2u_{-1}u_{0x} + 2u_0 u_{-1x} &
	u_{-1}^2 D + 2u_{-1} u_{-1x}  \\ u_{-1}^2 D + 2u_{-1} u_{-1x}
	& 0
	\endsmallmatrix \right)
	\left(\smallmatrix
	0 & \fr1{u_{-1}^2} D^{-1} \fr1{u_{-1}^2}  \\
	\fr1{u_{-1}^2} D^{-1} \fr1{u_{-1}}  &
	-\fr1{u_{-1}^2} D^{-1} \fr{u_0}{u_{-1}^2} -
	\fr{u_0}{u_{-1}^2} D^{-1} \fr1{u_{-1}^2}
	\endsmallmatrix\right)
	\\
	& \qquad\qquad \left(\smallmatrix
	2u_0 u_{-1}D & u_{-1}^2 D  \\ u_{-1}^2 D & 0
	\endsmallmatrix\right)
	\\
	& = \left(\smallmatrix 4u_0 & u_0^2 + 4u_{-1} \\
	u_0^2 + 4u_{-1} & 4u_0 u_{-1} \endsmallmatrix\right)
	\; D + \left(\smallmatrix 2 & 2u_0 \\ 0 & 2u_{-1} 
		\endsmallmatrix\right)
	\;u_{0x} + \left(\smallmatrix 0 & 2 \\ 2 & 2 u_0 \endsmallmatrix
	\right)\;
	u_{-1x}\;.
\endalign
$$
Again, this corresponds to a bracket of hydrodynamic type and
the non-degeneracy of the metric is characterized by the same
condition in (6.17).

\remark
{Remark 6.22}  (a) \ Alternatively, on the symplectic leaves of
the first structure defined by the conditions $\int u_0(x) dx = 
\const$, $\int u_{-1}(x)dx = \const$, $B_{-1}$ is invertible 
and therefore we can compute the recursion operator
$$
	\RC = B_0 B_{-1}^{-1} = \pmatrix
	u_0 + u_{0x} D^{-1} & 2  \\	
	2u_{-1} + u_{-1x} D^{-1} & u_0
	\endpmatrix
\tag6.23
$$
>From this, we can check that $B_1 = \RC B_0$.

(b) \ In principle, one can compute all higher structures 
explicitly by applying Dirac reduction to $\face_{(n)}$
or by using the recursion operator $\RC$, but the calculations are
quite formidable and we do not know if there exists an efficient
way to do this.
\endremark

\subhead
$2^{\circ}$ \ The dispersionless Toda lattice hierachy
\endsubhead

Let $A$ be the algebra introduced in Example 1$^{\circ}$, but
now we equip it with the following Lie bracket
$$
	[u,v]_0 = \lambda \left( \fr{\partial u}{\partial\lambda}
	\;\fr{\partial v}{\partial x} - \fr{\partial u}{\partial x}
	\fr{\partial v}{\partial \lambda}\right)\;, \qquad
	u,v \in A\,.
\tag6.24
$$
Then $(A, [\cdot,\cdot]_0)$ is also a Poisson algebra.  The
dispersionless Toda lattice hierachy is defined by the Lax
equations
$$
	\fr{dL}{dt} = \left[ \Pi_{\frak{k}} (L^n),L\right]_0
	= - \left[ \Pi_{\frak{l}} (L^n), L\right]_0\,,
	\quad n = 1,2,\dots
\tag6.25
$$
where the Lax operator $L$ is an element of the manifold 
$$
	\moda = \{ L \in A \spaces L(x,\lambda)
	= u_1 (x) \lambda + u_0(x) + u_1(x) \lambda^{-1}\}
\tag6.26
$$
and $\Pi_{\frak{k}}, \Pi_{\frak{l}}$ are the projection operators
relative to the direct sum decomposition
$$
	A = \frak{k} \oplus \frak{l}
\tag6.27
$$
into subalgebras
$$
	\frak{k}  = \left\{ u \in A \spaces u(x,\lambda) = \sum_{i > 0}
		u_i(x)(\lambda^i - \lambda^{-i})\right\}
\tag6.28a
$$
$$
	\frak{l}  = \left\{ u \in A \spaces u(x,\lambda) = \sum_{i \leq 0}
		u_i(x)\lambda^i\right\}\;.
\tag6.28a
$$
When $n = 1$, the corresponding Lax equation
$$
	\fr{dL}{dt} = \left[\Pi_{\frak{k}} (L),L\right]_0 \Longleftrightarrow
	\left\{
	\aligned
		u_{0t} & = 4u_1 u_{1x}   \\	
		u_{1t} & = u_1 u_{0x}
	\endaligned
	\right.
\tag6.29
$$
These are the dispersionless Toda lattice equations and can be
obtained from the periodic Toda lattice ODE system
$$
	\fr{da_k}{dt} = 2(b_k^2 - b_{k-1}^2)\,, \quad
	\fr{db_k}{dt} = b_k (a_{k+1} - a_k)
\tag6.30
$$
by taking a continuum (or long wave) limit.

The Poisson algebra $(A, [\cdot,\cdot]_0)$ also has all the
ingredients needed for the construction in Theorem 3.2.  In
this case, the invariant trace is of the form
$$
	\tree_0 u = \int u_0(x) dx\,, \qquad u \in A
\tag6.31
$$
which gives rise to the non-degenerate ad-invariant pairing
$(\cdot,\cdot)_0$:
$$
	(u,v)_0 = \tree_0(uv)\,, \qquad u,v \in A\,.
\tag6.32
$$
As the $r$-matrix for the equations in (6.25) is given by
$$
	R = \Pi_{\frak{k}} - \Pi_{\frak{l}}\,,
\tag6.33
$$
it follows from (3.3) that the Hamiltonian vector field 
generated by $H$ in the structure $\face_{(n)}$ 
is of the form
$$
\align
	X_H^{(n)}(u) & = \left[ \Pi_{\frak{k}} (u^{n+1} dH(u)),u\right]_0
	- u^{n+1} \Pi_{\frak{l}}^* ([dH(u),u]_0)
	\tag6.34
	\\
	& = u^{n+1} \Pi_{\frak{k}}^* ([dH(u),u]_0) - \left[ \Pi_{\frak{l}}
	(u^{n+1} dH(u)),u\right]_0\,.
\endalign
$$
Using this formula, we can now check that $\moda$
is a Poisson submanifold of $(A, \face_{(n)})$ only
for $n = -1,0$, and $1$.  Accordingly, the induced structures 
on $\moda$ provide the first, second and third Poisson
structures for the equations in the dispersionless Toda
lattice hierachy.  Using $u = (u_0,u_1)$ as coordinates
on $\moda$, the Hamiltonian operator of the first structure
is given explicitly by
$$
	B_{-1}(u) = \pmatrix 0 & u_1 \\ u_1 & 0 \endpmatrix
	D + \pmatrix 0 & u_{1x}  \\ 0 & 0 \endpmatrix\;.
\tag6.35
$$
Clearly, the associated Hamiltonian vector fields preserve the
sign of $u_1$.  Therefore, $B_{-1}(u)$ restricts to a structure
on
$$
	\moda^+ = \{ L \in A \spaces L(x,\lambda) = u_1(x)\lambda
	+ u_0(x) + u_1(x) \lambda^{-1}\,, \quad u_1(x) > 0 \}
\tag6.36
$$
whose symplectic leaves are the level sets of the Casimirs
$\int u_0(x) dx$, $\int \ln u_1(x) dx$.  Finally, we note that
$B_{-1}(u)$ is obviously of hydrodynamic type and the corrresponding
metric is non-degenerate on $\moda^+$ as det $(g^{ij}) = - u_1^2$.

\remark
{Remark 3.37}  Note that the equations in the dispersionless
Toda lattice hierachy are (strictly) hyperbolic in $\moda^+$
and we can take
$$
	w_1(u) = u_0 - 2u_1\,, \qquad w_2(u) = u_0 + 2u_1
\tag6.38
$$
as the Riemann invariants.  We shall not give the proof as the
reader can easily supply the details.
\endremark

As for the second and third Poisson structures on $\moda$, direct
calculation shows the corresponding Hamiltonian operators have
the form
$$
	B_0(u) = \pmatrix 4u_1^2 & u_0 u_1 \\ u_0 u_1 & u_1^2
	\endpmatrix D + \pmatrix 0 & 0 \\ u_1 & 0 \endpmatrix
	u_{0x} + \pmatrix 4u_1 & u_0  \\ 0 & u_1 \endpmatrix
	u_{1x}
\tag6.39
$$
$$
\align
	& B_1(u) =  \tag6.40   \\
	& \pmatrix 8u_0 u_1^2 & 4u_1^3 + u_0^2 u_1  \\
	4u_1^3 + u_0^2 u_1 & 2u_0 u_1^2 \endpmatrix D +
	\pmatrix 4u_1^2 & 0 \\ 2u_0 u_1 & u_1^2 \endpmatrix
	u_{0x} + \pmatrix 8u_0 u_1 & u_0^2 + 8u_1^2  \\
	4u_1^2 & 2u_0 u_1 \endpmatrix u_{1x}\;.
\endalign
$$
These structures also restrict to $\moda^+$, and are obviously
of hydrodynamic type.  But in contrast to the first structure,
the metrics associated with $B_0(u)$ and $B_1(u)$ are non-degenerate
only on a subset of $\moda^+$, characterized by the condition
$$
	w_1(u) w_2(u) \neq 0\,,
\tag6.41
$$
where $w_1(u),w_2(u)$ are the Riemann invariants in (6.38).

In order to compute the higher structures, we have to invoke
Dirac reduction, as in the last example.  Here, we shall do this
for the fourth structure as it presents new features which are
also shared by all higher structures.  First of all, we check
that for $L \in \moda^+$, we have $X_H^{(2)}(L) \in Im \Pi_{\frak{l}}^*$,
and the highest order term in $\lambda$ is $\lambda^2$.  Then we
write down the operator which gives $X_H^{(2)} (L)$ using the 
coordinates $u = (u_0,u_1,u_2)$ on the submanifold where
$X_H^{(2)} (L)$ lies:
$$
\left(
\smallmatrix
	(12 u_1^4 + 12 u_0^2 u_1^2)D + 6 (u_1^4 + u_0^2 u_1^2)_x
	&
	(u_0^3 u_1 + 12u_0 u_1^3)D + 6 u_1^3 u_{0x} + 24u_0 u_1^2
	u_{1x} + u_0^3 u_{1x} 
	& 	 2u_1^4 D + 8u_1^3 u_{1x}  
	\\	
	(u_0^3 u_1 + 12u_0 u_1^3) D + u_1(u_0^3 + 6u_0 u_1^2)_x 
	& 	(4u_1^4 + 3u_0^2 u_1^2)D + 3u_0 u_1^2 u_{0x} + (8u_1^3 +
	3u_0^2 u_1)u_{1x} & u_1^3 u_{0x}     
	\\
	2u_1^4 D &
	- u_1^3 u_{0x}  & - u_1^4 D - 2u_1^3 u_{1x}
\endsmallmatrix  \right)
\tag6.42
$$
Finally, we invoke Dirac reduction with constraint $u_2 \equiv 0$
to compute the Hamiltonian operator of the fourth structure, and
the result is 
$$
\align
	B_2(u) & = \pmatrix 16 u_1^4 + 12 u_0^2 u_1^2 &
	12 u_0 u_1^3 + u_0^3 u_1  \\ 12 u_0 u_1^3 + u_0^3 u_1 
	& 4u_1^4 + 3u_0^2 u_1^2 \endpmatrix D + \pmatrix
	12 u_0 u_1^2 & 4u_1^3  \\ 3u_0^2 u_1 + 8 u_1^3 & 3u_0 u_1^2
	\endpmatrix u_{0x}  \tag6.43
	\\
	& + \pmatrix 32 u_1^3 + 12 u_0^2 u_1 & 24 u_0 u_1^2 + u_0^3
	\\ 12 u_0 u_1^2 & 8 u_1^3 + 3u_0^2 u_1 \endpmatrix u_{1x}  \\
	& - \pmatrix 16 u_1 u_{1x} D^{-1} u_1 u_{1x} & 4 u_1 u_{1x}
	D^{-1} u_1 u_{0x}  \\ 4 u_1 u_{0x} D^{-1} u_1 u_{1x} &
	u_1 u_{0x} D^{-1} u_1 u_{0x} \endpmatrix \;.
\endalign
$$
Thus, $B_2(u)$ has a nonlocal tail, and provides an example of a class
of nonlocal Hamiltonian operators of the form
$$
	B^{ij}(u) = g^{ij} D + b_k^{ij} u_x^k + \sum_{\alpha=1}^N
	(w^{\alpha})_k^i u_x^k D^{-1} (w^{\alpha})_{\ell}^j
	u_x^{\ell}\;.
\tag6.44
$$
In the case where $\det (g^{ij}) \neq 0$, we note that the
geometric root of such structures was discussed in \c{F} and
applied to the chromatography equations.  At this point, the
reader can check that the subset of $\moda^+$ where the metric
associated with $B_2(u)$ is on-degenerate is likewise defined
by (6.41).  Also, on the symplectic leaves of the first structure
where $\int u_0 (x) dx = \const$ and $\int \ln u_1(x) dx = \const$,
the recursion operator $\R = B_0 B_{-1}^{-1}$ exists and it is 
not hard to show that $B_1 = \RC b_0$ and $B_2 = \RC^2 B_0$.

\remark
{Remark 6.45}  In \c{DM}, the authors considered the dispersionless
Toda lattice equations with boundary conditions $u_1(0) = 0$,
$u_1(1) = 0$.  We remark that the multi-Hamiltonian formalism
of this problem can also be obtained in a similar fashion.  Indeed,
the only major change one has to make here is to replace the algebra
above by the algebra of Laurent polynomials in $\lambda$, having
the form $u(x,\lambda) = \sum_i u_i (x) \lambda^i$, where the
coefficients $u_i$ are smooth functions on $I = [0,1]$ satisfying
the additional conditions $u_j(0) = u_j(1) = 0$, $j \neq 0$.
Otherwise, everything goes through just the same as before.  In
particular, the formula for the Hamiltonian operators of the first 
four structures are still those given in (6.35), (6.39), (6.40) 
and (6.43).
\endremark

In the next two examples, we shall consider equations with
infinitely many field variables.  For simplicity of exposition,
we shall not get into reduction calculations here, only remark
that the number of constraints is still finite in each case.

\subhead
$3^{\circ}$ \ The dispersionless KP hierachy
\endsubhead

Let $A$ be the algebra of formal Laurent series in $\lambda$,
having the form
$$
	u(x,\lambda) = \sum_{i=-\infty}^{N(u)} u_i (x) \lambda^i\,,
\tag6.46
$$
where the coefficients $u_i$ are smooth functions on $S^1 = \R/\Z$.
Define
$$
	[u,v]_{-1} = \fr{\partial u}{\partial\lambda}\;
	\fr{\partial v}{\partial x} - \fr{\partial u}{\partial x}\;
	\fr{\partial v}{\partial \lambda}\;, \qquad u,v \in A\,,
\tag6.47
$$
then $(A[\cdot,\cdot]_{-1})$ is a Poisson algebra.  The (extended)
dispersionless KP (dKP) hierachy is defined by the equations
$$
	\fr{dL}{dt} = \left[ \Pi_{\geq 0} (L^n),L\right]_{-1}
	= - \left[ \Pi_{\leq-1} (L^n),L\right]_{-1}\,,
	\qquad n = 1,2,\dots,
\tag6.48
$$
where the Lax operator is an element of the (extended) dKP
manifold
$$
	\M_{dKP} = \left\{  L \in A \spaces L(x,\lambda) =
	\lambda + \sum_{i = -\infty}^0 u_i(x) \lambda^i\right\}\,,
\tag6.49
$$
and $\Pi_{\geq 0},\Pi_{\leq-1}$ are projection operators
relative to the decomposition
$$
	A = A_{\geq 0} \oplus A_{\leq - 1}
\tag6.50
$$
into subalgebras
$$
	A_{\geq 0} = \left\{ u \in A \spaces u(x,\lambda) =
	\sum_{i \geq 0} u_i(x)\lambda^i \right\}\;.
\tag6.51a
$$
$$
	A_{\leq -1} = \left\{ u \in A \spaces u(x,\lambda) =
	\sum_{i =-\infty}^{-1} u_i(x)\lambda^i \right\}\;.
\tag6.51b
$$
In the standard form of the dKP equations \c{TT}, the coefficient
$u_0 \equiv 0$, but we shall not get into reduction calculations
here.

For the Poisson algebra $(A,[\cdot,\cdot]_{-1})$, the invariant
trace is defined by
$$
	\tree_{-1} u = \int u_{-1}(x) dx \;, \qquad u \in A\,,
\tag6.52
$$
and we have the non-degenerate ad-invariant pairing $(\cdot,\cdot)_{-1}$:
$$
	(u,v)_{-1} = \tree_{-1}(uv)\;,\qquad u,v \in A\;.
\tag6.53
$$
So again we can invoke Theorem 3.2,using the $r$-matrix
$$
	R = \Pi_{\geq 0} - \Pi_{\leq -1}
\tag6.54	
$$
in this case to obtain the corresponding brackets $\{\cdot,\cdot\}_{(n)}$,
$n \geq -1$.  Here, it is easy to check that $\M_{dkP}$ is a Poisson
submanifold of $(A,\{\cdot,\cdot\}_{(n)})$ only for $n = -1,0$.
Therefore, the induced structures on $\M_{dKP}$ provide the 
first and second Hamiltonian structures for the equations in the
hierachy.  For the bracket $\{\cdot,\cdot\}_{(1)}$, the slightly
larger manifold $\left\{ u \in A \spaces u(x,\lambda) = \sum_{i=-\infty}^1
u_i(x) \lambda^i \right\}$ is a Poisson submanifold.  Hence the
third structure on $\M_{dKP}$ can be computed using Dirac reduction
with constraint $u_1 \equiv 1$.  We shall leave the details to the
interested reader.

\subhead
$4^{\circ}$ \ The dispersionless modified KP and the dispersionless
	Dym hierachy
\endsubhead

Let $(A, [\cdot,\cdot]_{-1})$ be the Poisson algebra in Example 
3$^{\circ}$, with the same invariant pairing $(\cdot,\cdot)_{-1}$.
Consider the decomposition
$$
	A = A_{\geq k} \oplus A_{\leq k-1}\;,\qquad k \geq 0
\tag6.55
$$
with associated projection operators $\Pi_{\geq k}$ and 
$\Pi_{\leq k-1}$, where
$$
	A_{\geq k} = \left\{ u \in A \spaces u(x,\lambda) =
	\sum_{i \geq k} u_i(x) \lambda^i \right\}\,,
\tag6.56a
$$
$$
	A_{\leq k-1} = \left\{ u \in A \spaces u(x,\lambda) =
	\sum_{i =-\infty}^{k-1} u_i(x)\lambda^i \right\}\;.
\tag6.56b
$$
Clearly, $A_{\geq k}$ is a subalgebra of $(A,[\cdot,\cdot]_{-1})$
for all $k$.  On the other hand, simple verification shows that
$A_{\leq k-1}$ is a subalgebra of $(A,[\cdot,\cdot]_{-1})$ only
for $k = 0,1,2$.  Therefore, among the direct sum decompositions 
in (6.55), only the three cases $k = 0,1$, and $2$ lead to
$r$-matrices, and the case $k = 0$ has already appeared in
Example 3$^{\circ}$.  We now consider the other two cases, 
with Lax equations
$$
	\fr{dL}{dt} = \left[ \Pi_{\geq k} (L^n),L\right]_{-1}
	= - \left[ \Pi_{\leq k-1} (L^n),L\right]_{-1}\,,
	\qquad n = 1,2,\dots,\,;\;k = 1,2,\,.
\tag6.57
$$
For $k = 1$ and $L \in \M_{dKP}$, the equations in (6.57)
constitute the dispersionless modified KP hierachy.  For $k = 2$,
we obtain the dispersionless Dym hierachy when the Lax operator
$L$ is from the submanifold 
$$
	\M_{dDym} = \left\{ L \in A \spaces  L(x,\lambda) =
	\sum_{i=-\infty}^1 u_i(x)\lambda^i \right\}\,.
\tag6.58
$$
These hierachies are the semi-classical limit of the modified KP and
the Dym hierachies in \c{ANPV,KO}.  For the dmKP hierachy, with
$r$-matrix given by $R = \Pi_{\geq 1} - \Pi_{\leq 0}$, the
manifold of Lax operators is a Poisson submanifold of the associated
brackets $\{\cdot,\cdot\}_{(n)}$ for $k = -1,0,1$.  Hence the induced
structures on $\M_{dKP}$ provide the first three Poisson structures
for the Hamiltonian description of dmKP.  The higher structures, on
the other hand, have to be computed using Dirac reduction.  For the
dispersionless Dym hierachy, the situation is even better, for in this
case the first five Poisson structures on $\M_{dDym}$ are obtained
from the brackets $\{\cdot,\cdot\}_{(n)}$ $(-1 \leq n \leq 3)$
associated with $R = \Pi_{\geq 2} - \Pi_{\leq 1}$ by simple
restriction.  Again, the passage from $\{\cdot,\cdot\}_{(n)}$ $(n \geq
4)$ to the higher structures require the application of Dirac
reduction.

\enddocument